\documentclass{elsart}
\usepackage{graphics}
\usepackage{epsfig}
\usepackage{amssymb}
\begin{document}
\begin{frontmatter}
\title{The dynamics of entanglement in  two-atom Tavis-Cummings model with non-degenerate two-photon transitions for tripartite entangled $W$-like states}
\author{{{\corauthref{cor} E.K. Bashkirov}}, M.S. Mastyugin}
\address{Department of General and Theoretical Physics, Samara State University, 1 Academican Pavlov St., Samara,  443011,  Russia}
\corauth[cor]{Corresponding author. Tel. +79608129601, Fax: 88463345417}
\ead{ mast12basket@rambler.ru}
\begin{abstract}
The influence of dipole-dipole interaction on the entanglement between two atoms with different initial  entangled $W$-like states in Tavis-Cummings model with degenerate two-photon transitions has been investigated. The results show that the entanglement between two atoms can be increased by means of dipole-dipole interaction and for some initial states the entanglement sudden death effect can be weakened.
\end{abstract}

\end{frontmatter}

Entanglement is a key resource which distinguishes quantum
information theory from classical one. It plays a central role in
quantum information, quantum computation and communication, and
quantum cryptography \cite{Nielsen}. In recent years, there has
been a considerable effort to characterize entanglement
properties qualitatively and quantitatively   and to apply them in
quantum information. A lot of schemes are proposed for
many-particle entanglement generation. Yu and Eberly \cite{Yu} investigated the evolution of entanglement of a bipartite system in which the entanglement between two particles coupled with two independent cavities became completely vanishing in a time. This effect is designated the entanglement sudden death. Further investigations of this effect has been made by various authors (see Refs. in \cite{Yu1}). Resently, Zhang and  Chen \cite{Zhang} discussed the influence of dipole-dipole interaction on entanglement between two atoms with non-degenerate two-photon transitions. The result showed that the introduction of dipole-dipole interaction weaken the sudden death effect. The authors  make the cavity initially in the vacuum state and the two-atoms in a pure Bell states. Enlightened by the above work, we consider the influence of dipole-dipole interaction between on entanglement dynamics of two atoms with different initial tripartite entanglement states of $W$-type for non-degenerate two photons Tavis-Cummings model. In \cite{Bash}
we have obtained an exact expression for the evolution operator for two dipole coupled  atoms with non-degenerate two-photons transitions. This result allows us to find the explicit form of the reduced two-atom density matrix for any initial state of the system under consideration. Note also that the influence of dipole-dipole interaction on the entanglement sudden death of atoms with W-like initial entangled states for the two-atom Tavis-Cummings model with one-photon transitions has been considered by Li et al. \cite{Li}.

Let us consider two identical two-level atoms resonantly interacting with two-mode quantum electromagnetic field in lossless cavity through non-degenerate atomic transitions taking into account the dipole interaction between atoms. In the interaction picture and RWA approximation the Hamiltonian of considered model can be given
$$
H_{I} = \hbar g \sum\limits_{i=1}^2 (a_1^{+}a_2^+ R_i^- + R_i^+ a_1 a_2 )
 + \hbar \Omega (R_1^+ R_2^- + R_2^+ R_1^- ),\eqno{(1)}
$$
where $a_j^+$ and  $a_j$ denote the creation and annihilation operators of the frequency $\omega_j$ quantization field ($j=1,2$), $R_i^{+}=|+\rangle_{ii}\langle-|,\> R_i^{+}=|-\rangle_{ii}\langle+| $ are the atomic operators with $|+\rangle_i$ and $|-\rangle_i$ being the excited and ground states of the $i$th atom ($i=1,2$), g is the coupling constant between atom and field and $\Omega$ is dipole-dipole coupling strength between atoms.

 The density operator for the atom-field system follows a unitary time evolution generated by the evolution operator
 $U(t) = \exp(-\imath H t/\hbar)$. In the two-atom basis, $\> \mid +, + \rangle, \,\mid +, - \rangle, \,\mid -, + \rangle, \,
 \mid -, - \rangle$, the analytical form of the evolution operator $U(t)$ for system with Hamiltonian (1) is given by
  \cite{Bash}
\begin{center} $\quad$
$$
U(t) = \left (
\begin{array}{cccc}\vspace{2mm}
U_{11} & U_{12} & U_{13} & U_{14}\\ \vspace{2mm}
U_{21} & U_{22} & U_{23} & U_{24}\\ \vspace{2mm}
U_{31} & U_{32} & U_{33} & U_{34}\\ \vspace{2mm}
U_{41} & U_{42} & U_{43} & U_{44}\\
\end{array} \right ).\eqno{(2)}$$
\end{center}
Here $$U_{11} = 1 + 2 a_1 a_2
\frac{A}{\lambda} a^+_1 a^+_2,\quad U_{14} =  2 a_1 a_2
\frac{A}{\lambda} a_1 a_2,
$$
\vspace{2mm}
$$
U_{41} =  2 a^+_1 a^+_2 \frac{A}{\lambda} a^+_1 a^+_2,\quad
U_{44} = 1 + 2 a^+_1 a^+_2 \frac{A}{\lambda} a_1 a_2,
$$
\vspace{2mm}
$$
U_{12} = U_{13} = a_1 a_2 \frac{B}{\theta},\quad U_{21} = U_{31} =
\frac{B}{\theta} a^+_1 a^+_2 ,$$
\vspace{2mm}
$$ U_{24}=U_{34} =
\frac{B}{\theta} a_1 a_2 ,\quad U_{42} = U_{43} = a^+_1 a^+_2
\frac{B}{\theta},
$$\vspace{2mm}
$$
U_{22} = U_{33} = $$ $$ = \frac{\exp\left [- \imath
\frac{g}{2}(\alpha+\theta)t \right]}{4\theta} \left \{ [1 -
\exp(\imath g \theta t)] \alpha + 2 \theta
\exp(\imath \frac{g}{2}(3 \alpha + \theta) t] + \theta [1 +
\exp(\imath g \theta t)]\right \},
$$\vspace{2mm}
$$
U_{23} = U_{32}= $$ $$ = \frac{\exp\left [- \imath
\frac{g}{2}(\alpha+\theta)t \right]}{4\theta} \left \{ [1 -
\exp(\imath g \theta t)] \alpha -  2 \theta
\exp(\imath \frac{g}{2}(3 \alpha + \theta) t] + \theta [1 +
\exp(\imath g \theta t)]\right \},
$$
where
 $$ A = \exp \left [ - \imath \frac{g\alpha}{2}  t\right ] \left
 \{ \cos \left (\frac{g\theta}{2} t \right ) + \imath
 \frac{\alpha}{\theta} \sin \left ( \frac{g\theta}{2} t \right )
 \right \} - 1, $$\vspace{2mm}
$$ B = \exp \left [ - \imath \frac{g}{2}(\alpha + \theta)  t \right ] \left
 [1 - \exp(\imath g \theta t)\right ] $$\vspace{2mm}
è
$$ \alpha = \frac{\Omega}{g},\quad \lambda = 2(a_1 a_2 a^+_1 a^+_2 + a^+_1 a^+_2 a_1
a_2), \quad \theta = \sqrt{8(a_1 a_2 a^+_1 a^+_2 + a^+_1 a^+_2 a_1
a_2)+ \alpha^2}.
$$

We choose the atoms and the cavity to be prepared in an entangled
$W$-like state at the initial time,
 $$ |\Psi(0)\rangle = a|+,-;0,0\rangle + b|-,+;0,0\rangle + c |-,-;1,1\rangle, \eqno{(3)}$$
  where  $|n_1,n_2\rangle$ is two-mode Fock field and
     $|a|^2 + |b|^2 + |c|^2 =1$.

  Because the state of the system follows a unitary time evolution, we can acquire the corresponding time evolution of the system at time $t$, according to Eqs.(2), (3)
  $$ |\Psi(t)\rangle = X_1 |+,-;0,0\rangle + X_2|-,+;0,0\rangle + X_3 |-,-;1,1\rangle, \eqno{(4)}$$
  where
  $$X_1= (U_{22})_{0,0} \> a + (U_{23})_{0,0} \> b + \>\frac{B_{0,0}}{\theta_{0,0}}\>c,$$
  $$X_2= (U_{23})_{0,0} \> a + (U_{23})_{0,0} \> b +  \>\frac{B_0}{\theta_{0,0}}\>c,$$
  $$X_1=   \>\frac{B_{0,0}}{\theta_{0,0}}\>a +  \frac{B_{0,0}}{\theta_{0,0}}\>b  + (1 + 2 \frac{A_{0,0}}{\lambda_{0,0}})\>c.$$
  Here we denote
  $S_{n_1,n_2} = \langle n_1|\langle n_2 | S | n_1 \rangle|n_1\rangle,$ where $S$ is an arbitrary operator acting on field variables.

The information about the entanglement of two atoms is contained in the reduced density matrix $\rho_A(t)$ for two atoms which can be obtained from Eq. (4) by tracing out the state of the cavity
$$\rho_A(t) = Tr_F \>\> \rho_{AF}(t), $$
where
$$\rho_{AF}(t) = |\Psi(t)\rangle \langle \Psi(t)|.$$
In the two-atom basis  $|+,+\rangle, |+,-\rangle, |-,+\rangle, |-,-\rangle $ the reduced density matrix  $\rho_A(t)$ can be expressed as
\begin{center} $$\rho_{A} =
 \left ( \begin{array}{cccc}
 0 & 0 & 0 &0\\
  0 & |X_1|^2 & X_1 X_2^* &0\\
   0 & X_2 X_1^* & |X_2|^2 &0\\
    0 & 0 & 0 & |X_3|^2
\end{array}\right ). $$
\end{center}
It can be shown that the concurrence associated with the reduced density matrix is given by
$$ C(\rho_A) = 2\>\> {\rm max} \{ 0, |X_1 X_2| \}. \eqno{(5)}$$
Fig. 1 shows the concurrence  evolution of two atoms (5) with different initial states and dipole-dipole coupling strength values. It can observed that there is no sudden death phenomena when the system is in this initial state and that the entanglement between two atoms can be strengthened evidently by introducing the dipole-dipole  interaction.

 \begin{figure}[h!]
\begin{center}
\begin{tabular}{c}
\mbox{(a)} \\
\includegraphics[scale=0.65]{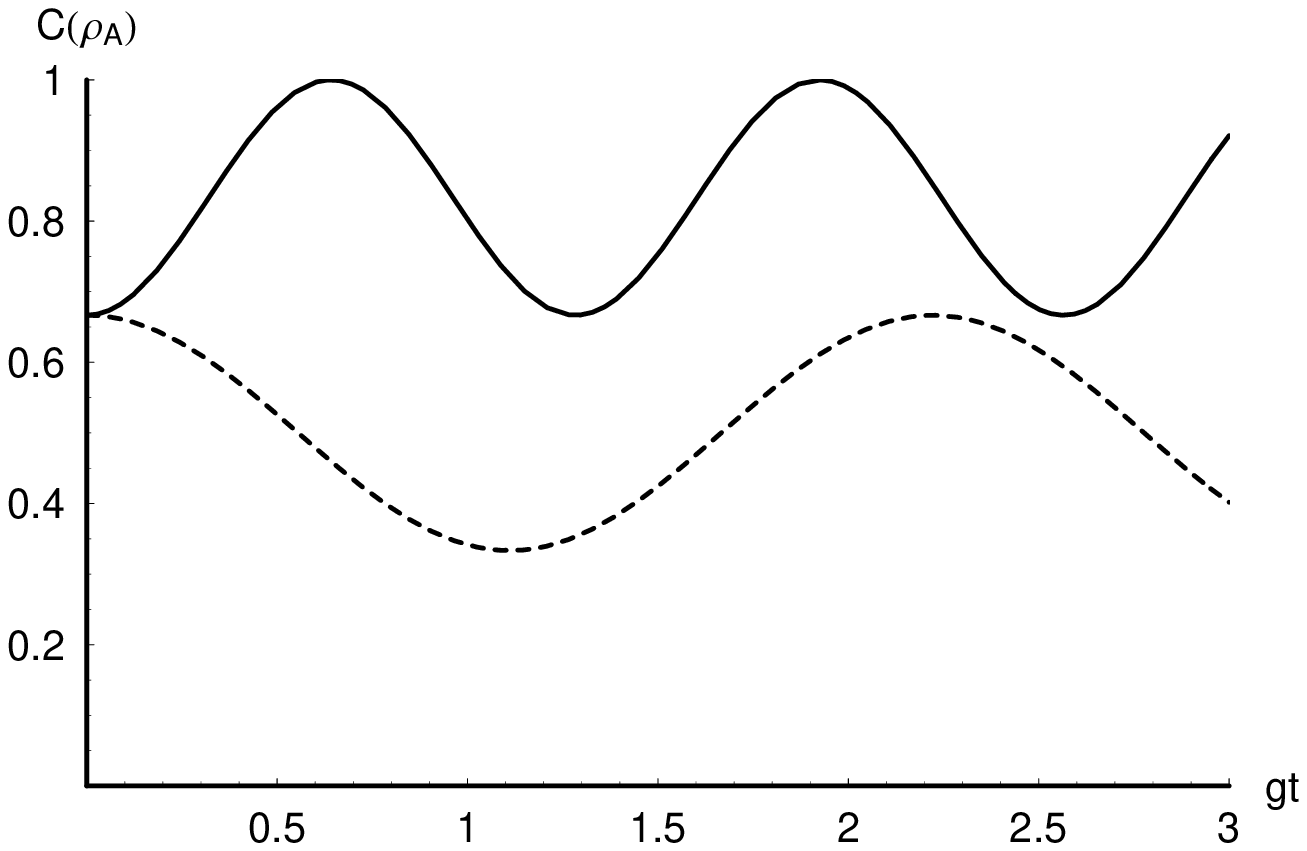}\\
 \mbox{(b)}\\
\includegraphics[scale=0.65]{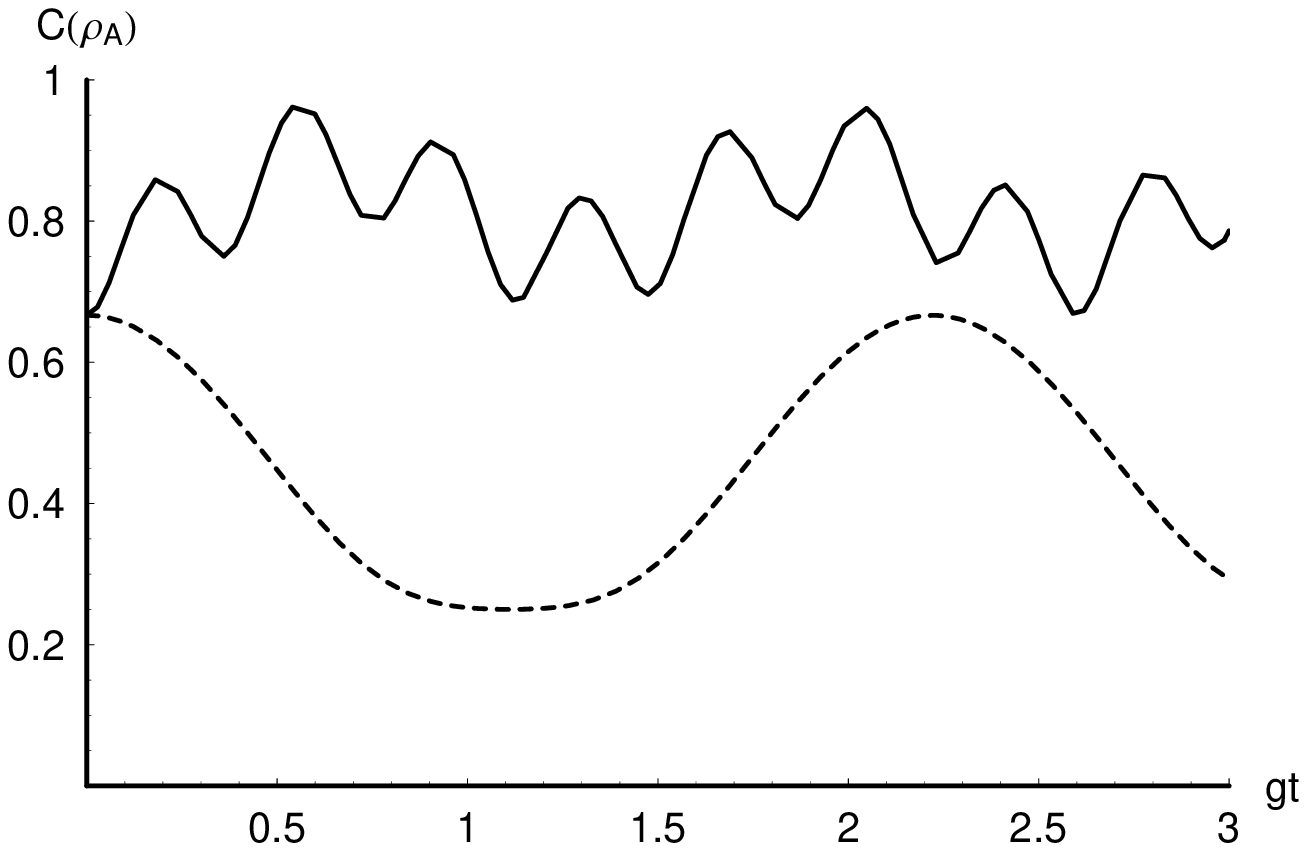}\\
\end{tabular}
\end{center}
\vspace{1mm}\begin{center}
\caption{The concurrence $C(\rho_A)$ for atom-atom entanglement when the whole system is initially prepared in  the  state (3) for (a) $a=b=c=1/\sqrt{3} $  and  (b)  $a= \sqrt{2/3},\> b=c=1/\sqrt{6} $  with  $\alpha  =6 $ (solid line) and $\alpha = 0 $ (dashed line).}
\end{center}
\end{figure}

Now we assume that the initial state for the total system is in a another form of $W$-like state,
$$ |\Psi(0)\rangle = a|+,+;0,0\rangle + b|+,-;1,1\rangle + c |-,+;1,1\rangle. \eqno{(6)}$$
In this case, the state of the total system governed by the Hamiltonian (1) at time t can be expressed in the standard basis
   $$ |\Psi(t)\rangle = X_1 |+,+;0,0\rangle + X2 |+,-;1,1\rangle + X_3|-,+;1,1 \rangle + X_4 |-,-;2,2\rangle,$$
where
  $$X_1 = (1 + 2 \frac{A_{1,1}}{\lambda_{1,1}})\> a + \>\frac{B_{1,1}}{\theta_{1,1}}\> b + \>\frac{B_{1,1}}{\theta_{1,1}}\>c,$$
  $$X_2=  \>\frac{B_{1,1}}{\theta_{1,1}}\> a + (U_{22})_{1,1} \> b + (U_{23})_{1,1}\>c,$$
 $$X_3=  \>\frac{B_{1,1}}{\theta_{1,1}}a + (U_{23})_{1,1} b + (U_{22})_{1,1} c,$$
  $$X_4=  4 \>\frac{A_{1,1}}{\lambda_{1,1}}\>a + 2\> \frac{B_{1,1}}{\theta_{1,1}}\> b  + 2\> \frac{B_{1,1}}{\theta_{1,1}}\> c.$$
The corresponding reduced density matrix have the form
\begin{center} $$\rho_{A} =
 \left ( \begin{array}{cccc}
 |X_1|^2 & 0 & 0 &0\\
  0 & |X_2|^2 & X_2 X_3^* &0\\
   0 & X_3 X_2^* & |X_3|^2 &0\\
    0 & 0 & 0 & |X_4|^2
\end{array}\right ). $$
\end{center}

\begin{figure}[h!]
\begin{center}
\begin{tabular}{c}
\mbox{(a)} \\
\includegraphics[scale=0.65]{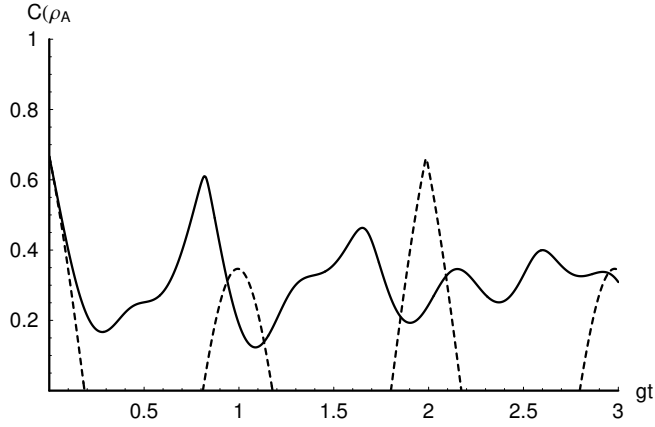} \\
\mbox{(b)}\\
\includegraphics[scale=0.65]{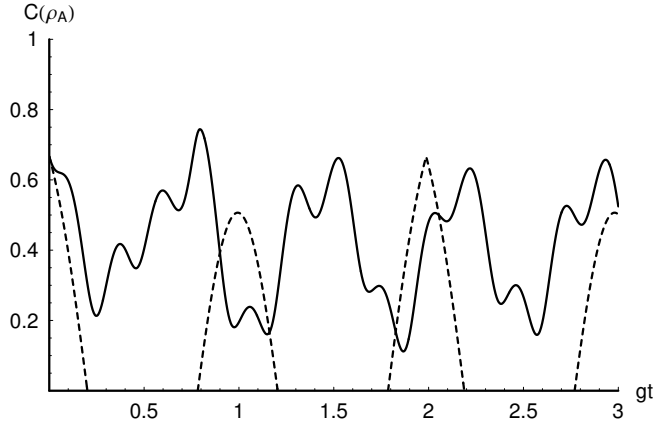}\\
\end{tabular}
\end{center}
\vspace{1mm}\begin{center}
\caption{The concurrence $C(\rho_A)$ for atom-atom entanglement when the whole system is initially prepared in the state (6) for (a)  $a=b=c=1/\sqrt{3} $  and (b)  $b= \sqrt{2/3},\> a=c=1/\sqrt{6} $ with $\alpha =6 $ (solid line) and $\alpha = 0 $ (dashed line).}
\end{center}
\end{figure}

The concurrence of this state can be obtained in the same way
$$ C(\rho_A) = 2 \>\> {\rm max} \{ 0, |X_2 X_3| - |X_1 X_4| \}. \eqno{(7)}$$
Fig. 2    shows the concurrence  evolution of two atoms (7) with different initial states and dipole-dipole coupling strength values. Unlike the previous case, one can see that entanglement can fall abruptly to zero and it will remain be zero for a period of time before entanglement recovers. Hence, the entanglement sudden death phenomena takes place in this case. The introduction of dipole-dipole interaction can completely eliminate the sudden death effect when the coefficients are chosen at $a=b=c=1/\sqrt{3}\>$ or $\> a=b=\sqrt{1/6},\> c=1/\sqrt{2/3}$. But this effect takes place only for large values of dipole-dipole strength (for example $\alpha$ should be more than  5 for the initial state with
$a=b=c=1/\sqrt{3}$).

In summary, we have investigated the entanglement character existed in non-degenerate two photons Tavis-Cummings model for different initial W-like states. The results demonstrate that for the two different initial states of the system, the entanglement evolution appears dramatically different. Therefore, the entanglement sudden death effect for considered model is very sensitive to the initial condition of the system. The introducing of dipole-dipole interaction can cause the entanglement to be higher and weaken the difference of entanglement evolvement between two kind differential initial state of system under consideration.
Such interaction also can weaken the sudden death effect and even completely eliminate this in some cases as well as in previous models \cite{Zhang}, \cite{Li}.

Work was carried out within the limits of the assignment of the Ministry of Education and Science of the Russian Federation  2.2459.2011.

\end{document}